\documentclass[10pt,conference]{IEEEtran}
\usepackage{cite}
\usepackage{amsmath,amssymb,amsfonts}
\usepackage{algorithmic}
\usepackage{graphicx}
\usepackage{textcomp}
\usepackage{xcolor}

\usepackage{listings}
\usepackage{diagbox}
\usepackage{fancybox}
\usepackage{makecell}

\usepackage{graphicx} 
\usepackage{subcaption} 
\graphicspath{ {fig/} }
\PassOptionsToPackage{hyphens}{url}\usepackage{url}
\usepackage{enumitem} 
\usepackage[normalem]{ulem} 
\usepackage{xspace}

\usepackage{amsmath,amssymb,mathtools}

\def\1{\mathbbm{1}}
\newcommand{\para}[1]{\left({#1}\right)}
\newcommand{\squarepara}[1]{\left[{#1}\right]}
\def\spara{\squarepara}

\renewcommand{\vec}[1]{\boldsymbol{#1}}
\newcommand{\mat}[1]{\boldsymbol{#1}}

\newcommand{\code}[1]{{\texttt{#1}}}
\usepackage{seqsplit}

\newcommand{\softmax}{\text{softmax}}
\newcommand{\Attn}{\text{Attn}}
\newcommand{\AttnTreeRel}{\text{Attn}_\text{TreeRel}}

\newcommand{\model}{\text{model}}

\newcommand{\abbr}[1]{\textrm{#1}}

\newcommand{\SrcSeq}{\abbr{SeqTrans}\xspace} 
\newcommand{\SrcRNN}{\abbr{SeqRNN}\xspace} 
\newcommand{\LeafSeq}{\abbr{LeafSeq}\xspace} 
\newcommand{\RootPath}{\abbr{PathTrans}\xspace} 
\newcommand{\DFS}{\abbr{TravTrans}\xspace} 
\newcommand{\TreeRel}{\abbr{TravTrans+}\xspace} 
\newcommand{\CSeq}{\abbr{Code2Seq}\xspace}
\newcommand{\ETH}{\abbr{Deep3}\xspace}
\newcommand{\PointerMixture}{\abbr{PointerMixture}\xspace}

\newcommand{\figref}[1]{Fig~\ref{#1}}
\newcommand{\secref}[1]{Sec~\ref{#1}}
\newcommand{\tabref}[1]{Table~\ref{#1}}

\definecolor{darkgreen}{RGB}{11,158,6}

\newcommand{\authornotesymbol}{{\textdagger}}
    
\begin{document}
\title{Code Prediction by Feeding Trees to Transformers}
\author{
    Seohyun Kim\textsuperscript{\authornotesymbol}\\
    \textit{Facebook Inc.}\\
    U.S.A. \\
    skim131@fb.com
    
    \and
    Jinman Zhao\textsuperscript{\authornotesymbol}\\
    \textit{University of Wisconsin-Madison}\\
    U.S.A. \\
    jz@cs.wisc.edu
    
    \and
    Yuchi Tian \\
    \textit{Columbia University}\\
    U.S.A. \\
    yuchi.tian@columbia.edu
    
    \and
    Satish Chandra \\
    \textit{Facebook Inc.}\\
    U.S.A. \\
    schandra@acm.org
}





\maketitle

\begingroup\renewcommand\thefootnote{\authornotesymbol}
\footnotetext{Both authors contributed equally to this research.}
\endgroup

\begin{abstract}
Code prediction, more specifically autocomplete, has become an essential feature in modern IDEs. Autocomplete is more effective when the desired next token is at (or close to) the top of the list of potential completions offered by the IDE at cursor position. This is where the strength of the underlying machine learning system that produces a ranked order of potential completions comes into play.

We advance the state-of-the-art in the accuracy of code prediction
(next token prediction) used in autocomplete systems.  Our work uses
Transformers as the base neural architecture.  We show 
that by making the Transformer architecture aware of the syntactic structure of code, we  increase the margin by which a Transformer-based system
outperforms previous systems.  With this, it outperforms the accuracy of several state-of-the-art next token prediction systems by margins ranging from 14\% to 18\%.

We present in the paper several ways of communicating the code structure to the Transformer, which is fundamentally built for processing sequence data. 
We provide a comprehensive experimental evaluation of our proposal, along with alternative design choices, on a standard Python dataset, as well as on Facebook internal Python corpus.  Our code and data preparation pipeline will be available in open source.

\end{abstract}


\section{Introduction}
\label{sec:introduction}

\subsection{Code Prediction}

\begin{figure}
    \centering
    \begin{subfigure}[t]{0.4\textwidth}
        \centering
        \includegraphics[width=0.9\textwidth]{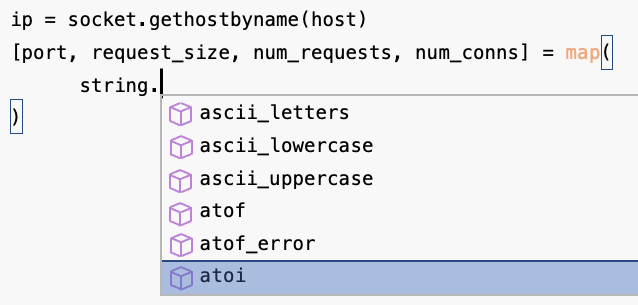}
        \caption{Type-based, alphabetical}
    \end{subfigure}
    \vspace{6pt}
    
    \begin{subfigure}[t]{0.4\textwidth}
        \centering
        \includegraphics[width=0.9\textwidth]{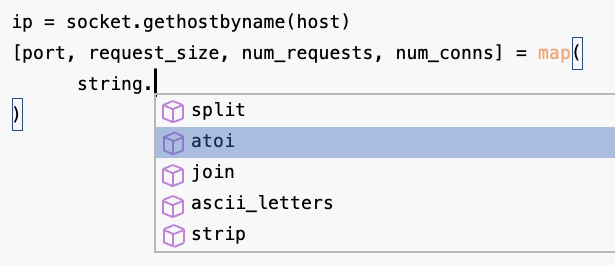}
        \caption{\SrcRNN}
    \end{subfigure}
    \vspace{6pt}

    \begin{subfigure}[t]{0.4\textwidth}
        \centering
        \includegraphics[width=0.9\textwidth]{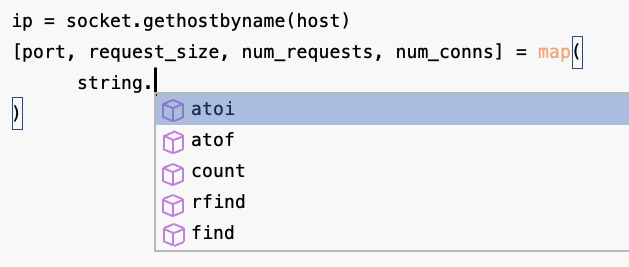}
        \caption{\DFS}
    \end{subfigure}
    \caption{Screenshots of ranked autocomplete predictions from three different models--Type-based alphabetical, \SrcRNN, and \DFS, as \textit{would} appear in an IDE.  A type-based autocomplete tool such as Jedi that sorts choices alphabetically ranks ``atoi'' low. 
    \SrcRNN, a RNN-based model predicts it as the second result.
    \DFS, a Transformer-based model predicts it as the first result. Fewer keystrokes are needed to choose the correct answer as we go from left to right.}
    \label{fig:jedi-atoi}
\end{figure}

The idea of code prediction in general is to predict some code element, given code surrounding the point of prediction. Code prediction is commonly used in an IDE for
\textit{autocomplete}, where based on the code already written up to the developer's cursor position, the IDE offers the most likely next tokens, perhaps as a drop down list to choose from
as shown in IDE views in Fig~\ref{fig:jedi-atoi}.  Other forms of code prediction could predict missing tokens at arbitrary code locations or predict larger units of code; in this paper we will concern ourselves with prediction of the immediate next token at the cursor position.

Consider the Python code fragment shown in Fig~\ref{fig:jedi-atoi}. 
Suppose a developer has written code up to \texttt{string} following by a dot. At this point, it will be helpful for the IDE to prompt the developer with attribute names that are \textit{likely} to follow, preferably, with \texttt{atoi} ranked at the top because in this case that is the correct next token.

Developers have come to rely on autocomplete in their IDEs for multiple reasons.  
First, and most obviously, it saves the effort of typing in the next token(s) in the IDE.  
For this reason alone, most modern IDEs come with at least some autocomplete facility 
for the languages they support.  Notice that a top-ranked suggestion is often selectable by hitting a tab---as would be the case in \figref{fig:jedi-atoi}(c)---whereas the lower ranked suggestions have to be selected by scrolling (\figref{fig:jedi-atoi}(a,b)), which is more effort.  Thus, providing the right completion at the top of the list, or if not, among the top few, is  important.

Second, autocomplete is also a powerful code discovery mechanism.  
For instance, a developer might not know the name of an API call they need off the top of their head, but is able to choose among the choices shown by an autocomplete tool.  Without assistance from IDE, 
the developer might need to change their mental context, go to Stack Overflow or some other web site,
and come back to the IDE.  
However, the code discovery assistance from autocomplete works only when a contextually appropriate code suggestion is offered among the \emph{top} choices in the list, because developers do not have the time to go through a comprehensive list of completions.





\subsection{Machine Learning for Code Prediction}

It is clear that effective autocomplete requires the intended next token to be predicted at the top of the list, or as close to the top as possible.  
Type-based autocomplete (e.g. Eclipse\footnote{\url{https://www.eclipse.org/pdt/help/html/working_with_code_assist.htm}} and Jedi\footnote{\url{https://github.com/davidhalter/jedi}}) returns a list of type-compatible names, but does naive ranking: alphabetically or with simple count-based statistics.
For instance, the autocomplete model used in the IDE (Jedi, which ranks 
type-compatible suggestions alphabetically) shown in Figure~\ref{fig:jedi-atoi}(a) ranks \texttt{atoi} fairly low.
The example shows why one of the earliest attempts of code prediction, using type-based methods, is not very effective.   Additionally, for dynamic languages, it is extremely difficult to gather an accurate type-compatible list of tokens that could occur in a context.

These limitations have motivated the use of machine learning for code prediction, as machine learning methods are able to base their predictions on the \emph{naturalness}~\cite{hindle2016naturalness} of code. 
Early approaches adapted $n$-gram language models on linearized source code tokens~\cite{allamanis2018survey, nguyen2013statistical-ngram}.
More recently,  deep neural networks have been applied to code prediction, surpassing $n$-gram models.
The most common neural technique for code prediction is Recurrent Neural Networks (RNNs)~\cite{liu2016neural-code-completion} and their variants~\cite{li2018code-rnn-attn, liu2020modeling-stack-lstm, karampatsis2020big-bpe, Svyatkovskiy_2019, raychev2014code-api,aye2020sequence}, where the code represented as a linear sequence is fed as input to the model. 
\figref{fig:jedi-atoi}(b) shows how an RNN model does better than a non-ML alphabetical ranking \figref{fig:jedi-atoi}(a) in showing the \textit{expected} item closer to the top.

Researchers have also investigated using the \textit{syntactic} structure of code for prediction, as opposed to seeing code as text: both using probabilistic graphical models (probabilistic context-free grammars~\cite{allamanis2014mining} and probabilistic higher-order grammars ~\cite{bielik2016phog,raychev2016learning-noisy,raychev2016probabilistic-deep3-eth-dt}), as well as using deep learning~\cite{alon2018codeseq}.

\begin{table*}
\centering
\begin{small}
    \centering
    \begin{tabular}{l|l|c|c|c|c|c|c|c|c|c|c|c|c}
    \multicolumn{1}{l|}{} & \textbf{Token value} & ip   & socket & get*Name & host & map  & string & atoi & sys  & argv & 2    & chain \\ \hline
    \textbf{Previous}    & \SrcRNN   & $>$10 & $>$10 & 3   & 2   & 7   & $>$10 & 2   & \textbf{1}   & \textbf{1}   & 3   & $>$10   \\
    \textbf{work}        & \ETH    & $>$10   & 9   & 4   & $>$10  &   $>$10  & $>$10    & $>$10    & 6    & \textbf{1}   & 7  &  $>$10  \\ 
    \multicolumn{1}{l|}{} & \CSeq & $>$10	& 2  & \textbf{1}	& 3 & $>$10	& $>$10 & 8	& \textbf{1} & \textbf{1}	& 3 & $>$10 \\ 
                        \hline
    \multicolumn{1}{l|}{} & \SrcSeq  & $>$10	& \textbf{1}	  & \textbf{1}	& 6   & $>$10	& $>$10 & \textbf{1}	& 10  & \textbf{1}	& \textbf{1}	  & $>$10   \\
    \textbf{Our}   & \RootPath 
                                    & 10	& 2	  & \textbf{1}	& \textbf{1}	  & 8	& \textbf{1}	  & 2	& \textbf{1}	  & \textbf{1}	& \textbf{1}	  & $>$10  \\
    \textbf{work}        & \DFS     & $>$10	& \textbf{1}	  & 5	& \textbf{1}	  & 4	& \textbf{1}	  & \textbf{1}	& \textbf{1}	  & \textbf{1}	& \textbf{1}	  & $>$10   
    \end{tabular}
\end{small}
    \caption{Ranks for the predictions for the leaf nodes listed in Fig~\ref{fig:exampleast}.  $>$10 means the model did not get the right answer in the top 10 results. }
    \label{tab:results_example}
\end{table*}

\subsection{Is this a solved problem?}

Although predictive models cannot be expected to be perfect, the accuracy of the current state-of-the-art methods leaves substantial margin to be improved.  For instance, a typical RNN-based method provides less than 37\% mean reciprocal rank (this equates to the correct answer being in the top $(37\%)^{-1} \approx 2.7$ results, \secref{sec:eval-metric}) on the py150 benchmark.
\textbf{Improving this metric is exactly the goal of this paper.} We report that our techniques are able to suggest the correct next tokens at ranks better --- showing correct result to the developer by 0.5 to 1 ranks higher --- than those achieved by previous methods (MRR increase of 14\% to 18\%.)

\begin{figure}
\centering

\lstset{basicstyle=\scriptsize\ttfamily,frame=single}

\begin{lstlisting}
...
ip = socket.gethostbyname (host)
[port, request_size, num_requests, num_conns] = map (
                string.atoi, sys.argv[2:]
)
chain = build_request_chain(num_requests, host, request_size)
...
\end{lstlisting}

    \caption{Running example of Python code. 
    The code snippet~\protect\footnotemark is from the py150 dataset~\cite{py150}.}
    \label{fig:examplecode}
\end{figure}

As a concrete example: Table~\ref{tab:results_example} shows the ranks of the various non-punctuation tokens to be predicted
for the code in Figure \ref{fig:examplecode}, using various recent methods, as well as for
our work.  Specifically, the rank of \texttt{atoi} is predicted at rank 1 \textit{only} by 
the new methods we propose in this paper.


\subsection{Feeding Trees to Transformers}

\footnotetext{\url{data/JeremyGrosser/supervisor/src/supervisor/medusa/test/test_11.py}}

To improve the accuracy of next token predicted, a number
of alternatives come to mind.  
First, we could strengthen 
the neural architecture alone.  For instance, researchers have
suggested adding \textit{attention} to RNNs
~\cite{iyer2016summarizing-lstm-attn,li2018code-rnn-attn} to compensate for loss of signal on long range dependence. 
Without long-range dependence a model would be making a (potentially sub-optimal) decision only on the most recent tokens.  Transformers handle long-range dependencies better.  
In the NLP community, \textit{Transformers} have achieved state-of-the-art results~\cite{devlin2018bert,dong2019unified-unilm,radford2019language-gpt2}, outperforming RNNs, 
for a variety of NLP tasks such as language modeling, question answering, and sentence entailment.  
For us, since training Transformers did not take much more resources that training RNNs (\tabref{tab:impldetails}), we decided to proceed with Transformers.

An orthogonal way to improve the accuracy is to enable
the machine learning system to ``see'' more code structure.   Raychev et al.~\cite{raychev2016probabilistic-deep3-eth-dt} had found that---for the code prediction problem---a non-neural but AST-aware engine could outperform RNNs.  In the same spirit, Alon et al.~\cite{alon2019code2vec} had found---for code summarization problem (though \emph{not} for code prediction in their paper)---that embedding the AST structure of code vastly outperformed purely sequence-based methods.

Inspired by the above, we explore how to leverage code structure while using Transformers on code.
This is not obvious; we cannot simply jam an AST into the Transformer, which is a sequence processing model. We explore two  models that represent two ways to capture the (partial) structure of an AST: one, based on decomposing the tree into paths (\RootPath), and the other, based on a tree traversal order (\DFS).
We also investigated a variant of \DFS that takes into account even more tree structure.


\subsection{Key Results}


\begin{table}
    \centering
    \begin{tabular}{|c||c|c|}
        \hline
                Model      &  Sequence  & AST \\  \hline
                \hline
        Transformer  & \textbf{\SrcSeq (54.9\%)} & \textbf{\DFS (58.0\%)} \\ 
                     &                   & \textbf{\RootPath (55.1\%)} \\ \hline
        Attention    &                   & \CSeq~\cite{alon2018codeseq} (43.7\%) \\ \hline
        RNN          &   \SrcRNN~\cite{hellendoorn2017deep} (36.6\%) & \\ \hline
        Decision tree &                  & \ETH~\cite{raychev2016probabilistic-deep3-eth-dt} (43.9\%) \\
        \hline

    \end{tabular}
    \caption{Overview of the models considered in this paper. Models in bold font are models from this work; \SrcSeq feeds source tokens in linear order to the Transformer.
    The numbers in parenthesis denote accuracy (see \secref{sec:evaluation}.) It is clear that the accuracy increases as models uses information from the AST (columns), and as more sophisticated neural architectures are used (rows). }
    \label{tab:modelchart}
\end{table}

Tab~\ref{tab:modelchart} compares the new models we introduce in this paper (in bold), with three state-of-the-art models from previous work (\secref{sec:backgroundtree} discusses these models.) We report results based on training and evaluating on the py150~\cite{py150} dataset, and for each, we report accuracy in mean reciprocal rank (MRR) as a percentage (see \secref{sec:implementation}).

Our best model \DFS, which communicates the tree structure to the Transformer, \textbf{significantly outperforms all previous models for code prediction,} with improvements in reciprocal ranks:
\begin{itemize}   
    \item from 43.9\% to 58.0\% when comparing a non-neural tree based model~\ETH\cite{raychev2016probabilistic-deep3-eth-dt} vs. \DFS;
    
    \item from 43.6\% to 58.0\% when comparing \CSeq ~\cite{alon2018codeseq} vs. \DFS;
    
    \item from 36.6\% to 54.9\% when comparing an RNN implementation \SrcRNN vs. \DFS\footnote{comparing \DFS against \SrcRNN is slightly different than comparing it against \ETH or \CSeq, hence the difference in MRR. We discuss in detail in \secref{sec:evaluation}.};
\end{itemize}   

We also evaluated our trained model on a dataset selected from a Python code repository \textit{internal} to a Facebook, and found the relative benefits of the Transformer models to be similar to those on the py150 dataset.  This indicates that the relative advantage of Transformer models carries over to other datasets. (\secref{sec:evaluation}, \tabref{tab:resultsint})

Deep learning models can be rather opaque as to their working.
To better understand whether \DFS is taking advantage of the tree structure, we 
employed saliency maps~\cite{simonyan2013deep-saliency} to examine where the model focuses its attention when making a prediction.  We found that indeed, the model tends to
focus on the most relevant parts of the tree, starting from the parent node (\secref{sec:inspection}).


\subsection{Contributions}
We move the state-of-the-art forward in accurate next token prediction, a capability increasingly
expected in modern IDEs.
\begin{itemize}
    \item We describe ways of using Transformers for the task of next token prediction, especially ways that \textit{profitably} communicate the syntactic structure of code. (\secref{sec:models}).
    Although there have been previous work in applying Transformers in the context of code (code summarization~\cite{great_iclr20}, code correction~\cite{harer2019tree-correction}, and code translation~\cite{shiv2019novel}), this paper is the among the first to explore and evaluate Transformers for code (next token) prediction.
    
    
    
    \item We present a systematic comparison of our proposed models with the most effective models from prior work\footnote{Incidentally, this paper is also the first to compare the three prior works on a common dataset.} that are applicable to next token prediction, on a widely available Python dataset \textit{py150}. The findings clearly indicate a 14\% to 18\% gain in accuracy with our best model relative to prior state-of-the-art. 
    
    \item  We provide a preliminary attribution study
    in an attempt to understand the prediction given by our best performing model, \DFS.  This study (\secref{sec:inspection}) indicates that \DFS indeed conditions its predictions on the most pertinent tokens in the context.  To our knowledge, this kind of model interpretability analysis for autocomplete is also a first.
    
\end{itemize}

Our overall conclusion is that Transformer based models over ASTs provide the best prediction power for autocomplete.  

\subsection{Outline} 
\secref{sec:models} provides background on the previous models that we use in our evaluation, along with a primer on Transformers.
\secref{sec:model-tree} explains the Transformer-based models of our own creation.
\secref{sec:implementation} describes our datasets and implementation. \secref{sec:evaluation} presents our quantitative results. \secref{sec:inspection} takes a closer look into why our models worked well (or did not). 
\secref{sec:threats} lists some threats to validity. 
\secref{sec:related-works} discusses prior related work in the field of code prediction and Transformers. We conclude the paper with our future work in \secref{sec:future-work}. 

\section{Background}
\label{sec:models}

\begin{figure}
\centering
\includegraphics[scale=0.3]{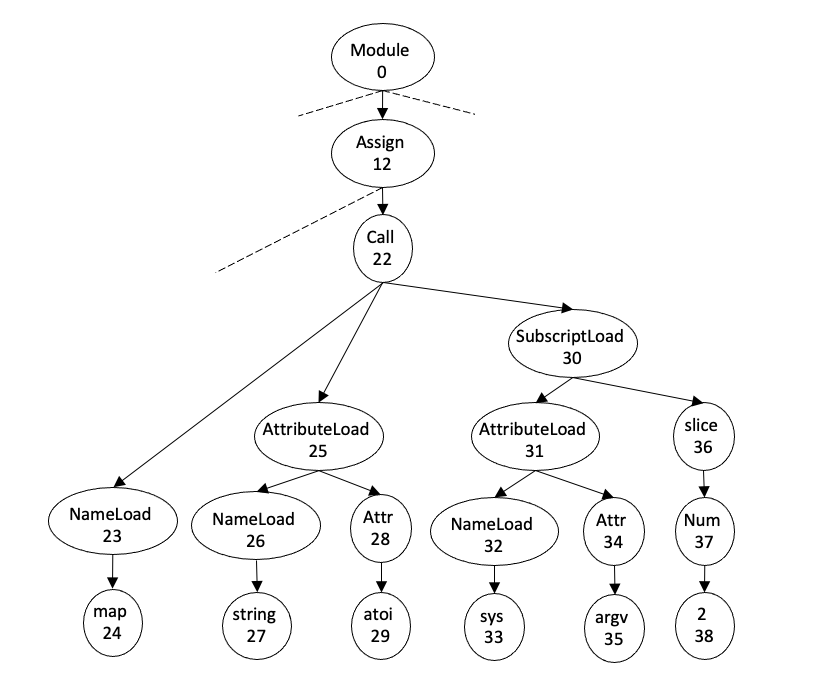}
\caption{Part of the AST for the example in Fig~\ref{fig:examplecode}. The leaf (terminal) nodes have values and the interior (non-terminal) nodes have types. }
\label{fig:exampleast}
\end{figure}

In this section, we define the code prediction task we examine in this work, followed by details of previous state-of-the-art methods of code prediction we use for comparison. We end the section with a brief introduction to the original Transformer model.  
We will refer to the nodes of the AST for Fig~\ref{fig:examplecode}, as shown in Fig~\ref{fig:exampleast}.

\subsection{Code Prediction Task}
\label{sec:code-prediction-task}
Code prediction task studied in this work is to predict the next code unit given the partial program up to the point of prediction. Let $p^*(unit \mid ctx)$ be the empirical distribution of code unit given the partial program context $ctx$. Our task is to learn to approximate $p^*$ using a machine learning model $M$. In our proposals, $M$ will be some Transformer $Trans$. 
The learned distribution can be viewed as 
\begin{align*}
    p(unit \mid ctx) = M(ctx; \mat{\theta}),
\end{align*}
where $\mat{\theta}$ represents the trainable parameters of the model.
We train the models by minimizing the KL-divergence between $p$ and $p^*$, or equivalently, minimizing the cross-entropy loss $l$ over all code prediction locations.

We explore several ways of representing partial program and predicting different kinds of code units. 
When using source code token as code unit and representing partial program as sequence of source code tokens (\SrcSeq), the problem aligns with the traditional notion of language modeling -- predicting the next token in a sequence given all previous tokens: $p(t_i \mid t_1, \dots, t_{i-1})$ where $t_i$ is the $i$-th token.

More interestingly, we explore various representations of a partial program to better utilize its AST information. The intuition is that the more we can utilize the syntactic information provided by the AST, the better we can predict the next token. The next section discusses three models used in previous work. 


\label{sec:backgroundtree}

\subsection{\textbf{\SrcRNN}}
\label{sec:srcrnn}

For next token prediction, a popular method is to feed the source sequence tokens into an RNN (or LSTM)~\cite{hellendoorn2017deep, Svyatkovskiy_2019, raychev2014code-api,aye2020sequence}.
An RNN embeds the input tokens into a vector: $\vec{x_t} = \mathit{emb}(\mathit{{w}_t})$, where $w_t$ is the source token seen at the $t$'th time step. 
The hidden state $\vec{h}_{t+1}$ at the $(t+1)$-th time step is
computed as $\vec{h}_{t+1} = rnn \para{\vec{x}_t, \vec{h}_{t} }$,
where $rnn$ is a trainable RNN unit. The last hidden state is then fed through a classification layer.

The pertinent point to note is that the hidden state $\vec{h_t}$ encodes the knowledge of not just the current token, but of last several (and theoretically all) previous tokens via the propagation of information in previous hidden states.  


In our experiments, we feed the source code tokens into an LSTM and call this model \SrcRNN.

\subsection{\textbf{\ETH}}

Raychev et al.~\cite{raychev2016probabilistic-deep3-eth-dt}
presented a system, Deep3, based on a learned decision tree combined with count-based probabilities at the leaves of the decision tree.  

\begin{figure}
\includegraphics[scale=0.25]{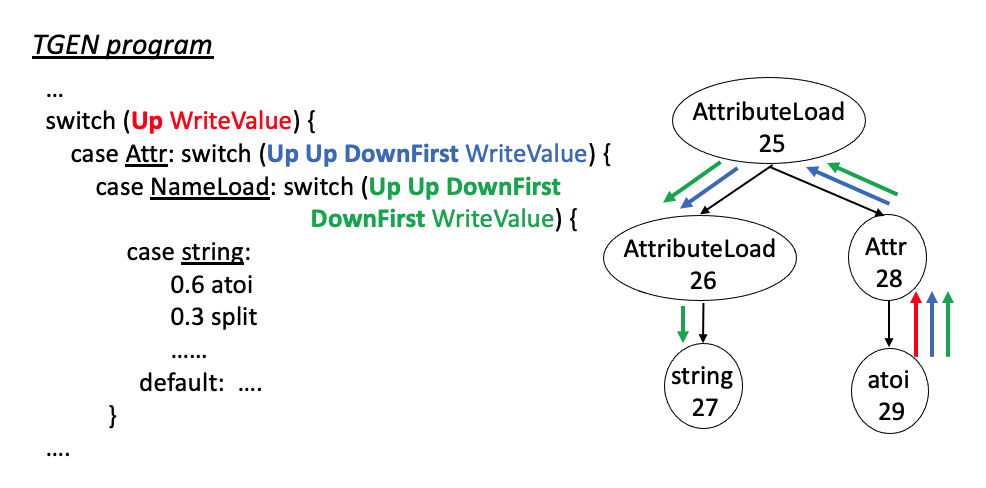}
\caption{Fragment of a TGEN program encoding a decision tree on the left (bold words are the steps that comprise a path), with the corresponding paths shown on the AST on the right.}
\label{fig:deep3}
\end{figure}

Fig~\ref{fig:deep3} shows part of a learned decision tree, written in the form of program in a specialized language called TGEN. 
Given an AST $t$ and a starting node $n$, a TGEN program walks certain paths in $t$ starting from $n$. For example, \texttt{Up WriteValue} (line 1) goes to the parent of $n$ and records the label. 
At the end of a TGEN program is a probability distribution for the possible values of the starting node. For example, starting with node 29, the TGEN program predicts ``atoi'' with 60\%, ``split'' with 30\%, etc. 

A TGEN program is learned---on a specific corpus---by a genetic search procedure that simultaneously selects paths and grows the decision tree from the training data, with an entropy minimization objective. 
In this paper, we use their pretrained model~\cite{phog} as well as their Python dataset~\cite{py150} for our experiments.

\subsection{\textbf{\CSeq}}

\begin{figure}
\includegraphics[scale=0.2]{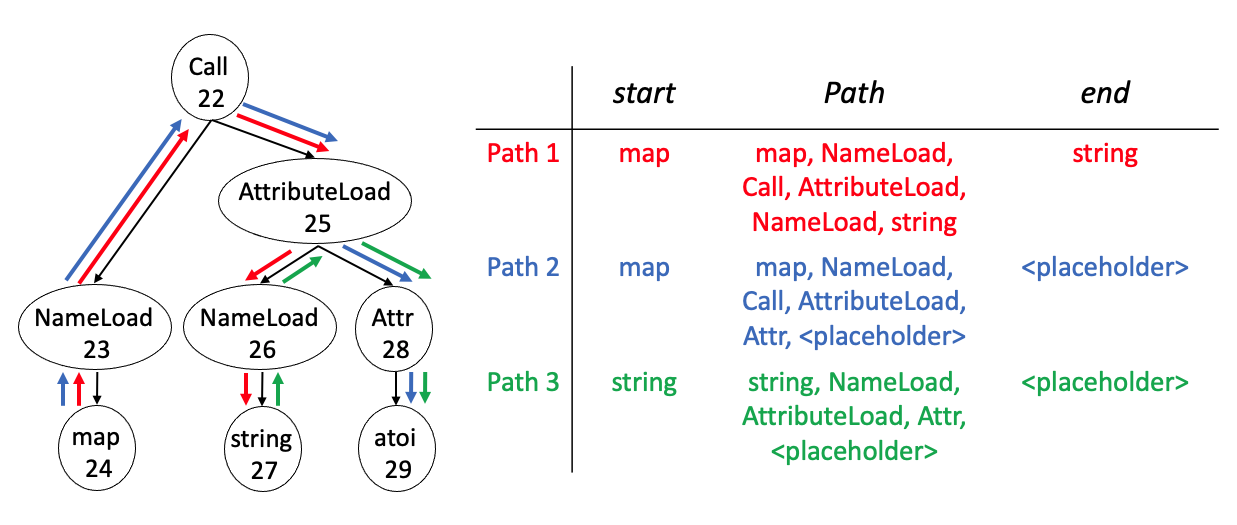}
\caption{Example of an input for \CSeq, which consists of leaf-to-leaf path representations given a partial AST. A path representation is made of tokenized starting tokens, path, and tokenized ending tokens. If the path ends with the target node (in this example, \texttt{atoi}), the value is replaced by \texttt{<placehholder>}.}
\label{fig:code2seqexample}
\end{figure}

\CSeq is a model by Alon et al. ~\cite{alon2018codeseq} that embeds code snippets by embedding AST paths in a neural network.

At a high-level, given an AST, \CSeq creates path representations for all leaf-to-leaf paths. 
For example, Fig~\ref{fig:code2seqexample} shows three leaf-to-leaf paths for nodes 22-29 from the full AST (Fig~\ref{fig:exampleast}).
For each path, a path representation is created with: 1. the starting leaf value, tokenized by snake and camel case, 2. the path itself, and 3. the ending leaf value, also tokenized. 
1 and 3 are embedded using LSTMs, and 2 is embedded using bi-directional LSTMs.
These three embeddings are concatenated and then fed through a feed forward network. Finally, all of the path represention embeddings in the AST are combined using a simple attention mechanism. 

In \CSeq, a decoder is then used to solve a code summarization task: 
given a method body, how well can \CSeq generate the correct method name? The training proposed in~\cite{alon2018codeseq} is not well suited for next token prediction. In code summarization, a set of leaf-to-leaf paths needs to be created one time for a method.  By contrast, in code prediction, a new set of leaf-to-leaf paths has to be created for each point of prediction.

For example, to predict \texttt{atoi} (node 29) in Fig~\ref{fig:exampleast}, we must first create an representative embedding for the partially completed AST up to node 29, using all leaf-to-leaf paths available up to node 29.
Paths that end in \texttt{atoi} are also used, with \texttt{atoi} replaced with a \textit{placeholder} token to prevent information leak (e.g. Paths 2 and 3 in Fig~\ref{fig:code2seqexample}). The representative embedding is then fed through a classification layer to generate predictions.\footnote{Note that this is different than the generative decoder that \CSeq uses.}

By treating each point of prediction as a separate data point (compared to a language model, where one sequence is considered one data point), the number of training data points, along with the effort to  create them makes \CSeq computationally very expensive.

\subsection{A Primer on Transformers}
\label{sec:trans-intro}

Here we present a brief introduction of Transformers. Readers familiar with Transformers can skip ahead to Section~\ref{sec:model-tree}.

Transformers belong to a class of deep neural networks that are designed for sequence processing.  
In Transformers, information from any previous location of the sequence can directly affect the encoding of the next token, through a mechanism called \textit{self-attention}, which helps greatly improve the connectivity in long sequences.  

To be precise, a Transformer is a stack of Attention blocks (\text{AttnBlk}) preceded by an input embedding layer ($\text{Emb}$) and followed by a classification layer ($\text{Clsfr}$), where $\text{AttnBlk}$ is repeated $n_{block}$ times.
\begin{align*}
    Trans(ctx) = \text{Clsfr}(\text{AttnBlk}( \dots (\text{AttnBlk}(\text{Emb}(ctx))) \dots ))
\end{align*}

\begin{figure}
    \centering
    \includegraphics[scale=0.4]{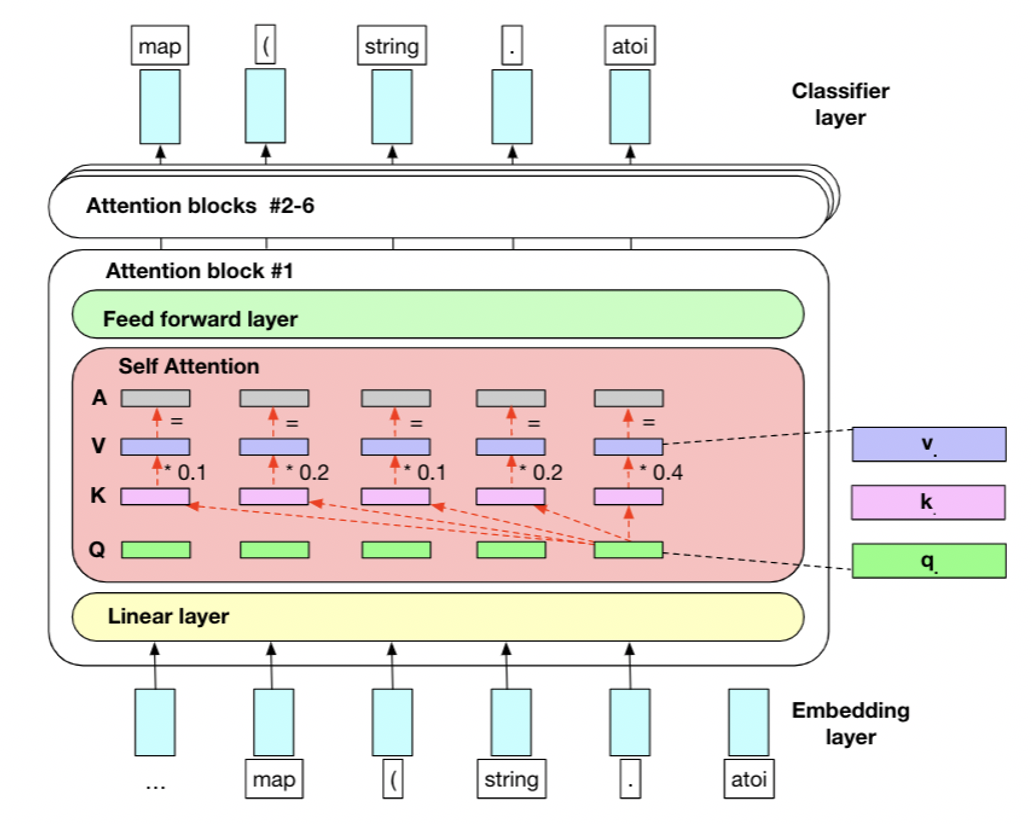}
    \caption{Schematic of a GPT2 Transformer. The self-attention layer is able to consider all tokens in the input up to the point of prediction. Here the self-attention box depicts the information flow when predicting next token after the "."; see Table~\ref{tab:QK_softmax} for where the numbers come from.} 
    \label{fig:gpt2}
\end{figure}

See \figref{fig:gpt2} for a schematic of a Transformer with
embedding layer, a stack of (here 6) attention blocks, and
finally a classification layer.

The self-attention layer---which constitutes the main part of an attention block---is the crux of the model. 
The intuition here is to attend to the elements in the input sequence in proportion of their relevance to the  location being predicted.
For example, take an example input token sequence [``map'', ``('', ``string'', ``.''], and the target next token being ``atoi.''
It is first fed through the initial embedding layer to give: $\mat{E} = \spara{e_\text{map}, e_\text{(}, e_\text{string}, e_\text{.}}$.
Then, we feed $\mat{E}$ to three fully-connected networks ($\mat{W_q}, \mat{W_k}, \mat{W_v}$) to create query, key, and value embeddings:
\begin{gather*}
  \mat{Q} = \mat{E} \mat{W_q},   \mat{K} = \mat{E} \mat{W_k},  \mat{V} = \mat{E} \mat{W_v},
\end{gather*}
\figref{fig:gpt2} depicts the vectors $\mat{Q}$ comprised of its elements $q_\text{map}, q_\text{(}, q_\text{string},  q_\text{.}$ (in green), and likewise for $\mat{K}$ and $\mat{V}$.

The self-attention then works by querying keys $k$ using queries $q$ and then using the result to summarize values $v$ through the attention function:
\begin{align*}
    \text{Attn}(\mat{Q}, \mat{K}, \mat{V}) = \softmax\para{\frac{\mat{Q}\mat{K}^\top}{\sqrt{d_k}}} \mat{V}
\end{align*}
where $d_k$ is the dimension of key vectors.
Here, $\mat{Q}\mat{K}^\top$ is used to determine which token relationships are the most important, resulting in a matrix of size $n \times n$, where $n$ is the length of the input sequence. Each row is then normalized and passed through a softmax layer.
~\tabref{tab:QK_softmax}  shows an example of the self-attention weights.
Looking at the last row, we can see that most of the attention is given to ``.'', meaning it has a greater factor in predicting the next token ``atoi''.
Note how the matrix is a lower triangular matrix: this is because self-attention cannot be applied to tokens that have not been seen before. After multiplying this matrix with the value vector, in our example, $\text{Attn}(\mat{Q}, \mat{K}, \mat{V}) = \spara{0.2 * v_\text{map}, 0.1 * v_\text{(}, 0.2 * v_\text{string}, 0.4 * v_\text{.}}$. 


Transformers also uses multiple heads of these self-attention blocks, called multi-headed attention, which enables the model to simultaneously consider different ways of attending to previous information within one block and also across other blocks. In our implementation, we omit positional encoding (see \secref{sec:impl}.)
For more details, please refer to \cite{vaswani2017attention} and \cite{radford2019language-gpt2}.

The next sections discuss various ways of feeding code fragments into this Transformer architecture.

\section{Our work: Transformer-based Models}
\label{sec:model-tree}

The question that interests us is: can Transformer-based models also benefit from syntactic structure, and if so, how can we communicate the syntactic structure to Transformer?

In this section, we first begin with a model \SrcSeq that uses a Transformer to take source code tokens as input. Then we introduce two other models, \RootPath and \DFS, that use more syntactic information obtained from the AST. 


\subsubsection{\textbf{\SrcSeq}}
Our first model is to apply a Transformer over source token sequences, which can be easily obtained by applying a tokenizer.
Here the input is the partial program represented as source token sequences and the output is a source code token. 
This is a straightforward application of the original Transformer design, and functions as a baseline for our later attempts that take on more AST information.
The model can be written as
\begin{align*}
    \vec{o} = Trans\para{\para{\vec{e}_{t}}_{t \in \mathit{src\_seq}}}
\end{align*}
where $\vec{o}$ is a distribution over all possible tokens, and $\vec{e}_{t}$ is the embedding for source token $t$ for every $t$ in the source token sequence.
As we show in the experiments, \SrcSeq turns out to be an already strong model as a direct comparison to the baseline RNN model \SrcRNN.

Since Transformers are originally designed as a sequential model, the challenge becomes finding ways to convey AST information to Transformers. 
In the next subsection, we will vary the inputs and the outputs to the Transformer, but the principles of operation will remain the same as in \SrcSeq.


\begin{table}
\centering
\begin{tabular}{|l|l|l|l|l|}  
\hline
... & map  & ( & string & . \\ \hline
map    & 0.9 &&& \\ 
(      & 0.6 & 0.1 && \\
string & 0.1   & 0.1  &  0.7  & \\
.      &  0.2     & 0.1  &  0.2  &  0.4 \\ \hline
\end{tabular}
\caption{Example matrix for the numerical self-attention scores after taking the softmax over the normalized values of $\mat{Q}\mat{K}^\intercal$. For example, the entry against
$(\text{string},\text{map})$ is obtained by multiplying 
$q_\text{string}$ with $k_\text{map}$, after softmax and normalization.
Note that the rows listed here do not sum up to exactly 1 since there are previous tokens in the input sequence that are not shown in this matrix.}
\label{tab:QK_softmax}
\end{table}

\begin{figure}
\includegraphics[scale=0.22]{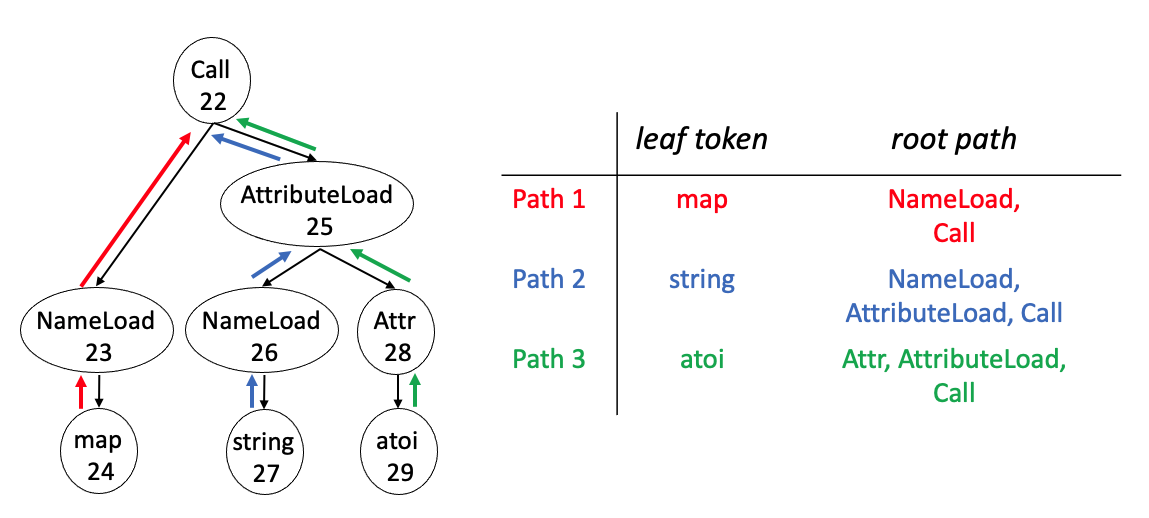}
\caption{Example of an input to the \RootPath model. It takes all leaf nodes, along with its path to root, to make the next leaf token prediction. The root paths are embedded using an LSTM, and the leaf tokens are embedded using an embedding layer. These two embeddings are added together to create a path representation embedding, which is then fed to the Transformer as shown in Fig~\ref{fig:gpt2}.
The classification layer of the Transformer outputs leaf tokens.}

\label{fig:rootpathexample}
\end{figure} 

\subsubsection{\textbf{\RootPath}}
\RootPath enhances \SrcSeq by exposing tree structure to the Transformer via root-paths. 
A root-path is the path from the leaf node $t$ to the root of the AST by traversing up its ancestors, recording all the nodes along with it,
and thus a sequence of internal AST nodes.
Fig~\ref{fig:rootpathexample} shows an example of an input datapoint for predicting node 29 of Fig~\ref{fig:exampleast}.


The root-paths are first fed into a an LSTM\footnote{We could have used a Transformer to embed the path sequences in lieu of an LSTM, but since the path sequences are short (capped at 13 tokens) enough for LSTMs to perform adequately well, we decided to use an LSTM. See \secref{sec:impl} for details.} added with the embedding of the leaf node, and is fed through a Transformer:
\begin{align*}
    \vec{o} = Trans\para{\para{\vec{e}_{t} + \vec{p}_{t}}_{t \in \mathit{leaf\_seq}}}
\end{align*}
where $\vec{o}$ is a distribution over all possible leaf nodes, and $\vec{e}_{t}$ is the embedding for AST node $t$ and $\vec{p}_t = \text{LSTM}\para{\para{\vec{e}_u}_{u \in Rootpath(t)}}$ is the summarized representation of root-path from $t$ for every leaf node $t$ in the leaf sequence $\mathit{leaf\_seq}$.
\footnote{$+$ is used here following the convention of Transformer computations and to keep the embedding dimension the same for every component.}
The hope here is that the root-paths captures the local syntactical information and thus can help the prediction.

Since the points of prediction are the leaf nodes of the AST, the loss is taken over only the leaf AST nodes, and it predicts only leaf tokens. 

\subsubsection{\textbf{\DFS}}
As a Transformer naturally only takes a sequence as input, we provide the AST nodes as a sequence in pre-order traversal, or a depth-first-search (DFS) order. For Fig~\ref{fig:exampleast}, for node 29, the previous nodes in DFS order would be:
[..., ``Call'', ``NameLoad'', ``map'', ``AttributeLoad'', ``NameLoad'', ``string'', ``Attr''].

The \DFS model can be written as:
\begin{align*}
    \vec{o} = Trans\para{\para{\vec{e}_{t}}_{t \in \mathit{AST\_seq}}}
\end{align*}
where $\vec{o}$ is a distribution over all possible tokens, and $\vec{e}_{t}$ is the embedding for AST token $t$ for every $t$ in the partial program represented as a AST token sequence $\mathit{AST\_seq}$ in DFS order.



\subsubsection{Capturing even more AST structure?}

\DFS presents the tree nodes in a pre-determined order, but still does not retain detailed structural relationship between nodes.  For example, consider the sequence of nodes 26 - 28 in \figref{fig:exampleast}. This would be represented as [``NameLoad'', ``string'', ``attr''], the three nodes appearing consecutively in DFS order. Looking at the AST, we can see that the relations between (``NameLoad'' \& ``string'', and ``string'' \& ``attr'') are actually quite different: ``NameLoad'' is one node up from ``string'', while ``string'' is two nodes up and one node down from ``attr''.

We create a new model, \TreeRel that enhances \DFS by capturing these richer path-based relations. 
Similarly to Hellendoorn et al.~\cite{hellendoorn2020global-relational}, we enhance the self attention block of the Transformer with a matrix $\mat{R}$ that captures the (unique) path needed to reach from $a$ to $b$. This path is represented abstractly only in terms of up and down moves: 
\begin{align*}
  \mathit{UDpath}(a, b) = U^i D^j
\end{align*}
where $i$, and $j$ are the number of up and down nodes, respectively, node $a$ has to travel to reach node $b$. For example, \textit{UDpath}(29, 27) = $U^2 D^2$ for node 29 and 27 in Fig~\ref{fig:exampleast}.
$\mat{R}$ is introduced to the Transformer by replacing the $\Attn$ function with the following $\AttnTreeRel$ function. 
\begin{equation*}
  \AttnTreeRel(\mat{Q},\mat{K},\mat{V},\mat{R}) = \softmax\para{\frac{\mat{R} \odot (\mat{Q} \mat{K}^\top)}{\sqrt{d_k}}} \mat{V}
\end{equation*}
where $\odot$ is element-wise product. This provides a way for the self attention to consider the previous tokens, taking into account the AST relationship between pairs of nodes as well.

\section{Implementation and Datasets}
\label{sec:implementation}

\subsection{Dataset}
We train our models using the py150 dataset used in \cite{raychev2016probabilistic-deep3-eth-dt}.
The dataset consists of 150k Python 2 source code files from GitHub repositories, along with their parsed ASTs, split into 100k for training and 50k for evaluation.
In this work, we slightly modify the AST to ensure that the internal nodes only carry syntactic types and the leaf nodes only carry token values.
To incorporate large trees (greater than 1000 nodes, which is the limit we chose for transformer window), we deploy a technique adopted by \cite{alrfou2018characterlevel}, which slices a large tree into shorter segments with a sliding window to maintain part of the previous context.

We evaluate our models on two evaluation datasets:
\begin{itemize}
    \item \textbf{py150}: We use the evaluation dataset used in \cite{raychev2016probabilistic-deep3-eth-dt}, which consists of 50k Python ASTs. After the modifications mentioned above, there are 16,003,628 leaf nodes.
    \item \textbf{internal}: We also created an evaluation dataset consisting of 5000 Python files from a code repository internal to Facebook. With this dataset, we can evaluate how our trained model can generalize to a different dataset, even if the code comes from disjoint projects. After the modifications, there are 1,669,085 leaf nodes.
\end{itemize}

\subsection{Implementation}
\label{sec:impl}

\paragraph{Transformers} For the models that use Transformers (\SrcSeq, \RootPath, \DFS, \TreeRel), we adapt the Pytorch implementation~\footnote{\url{https://github.com/graykode/gpt-2-Pytorch.}}
of GPT-2 small~\cite{radford2019language-gpt2}.
We use six Transformer blocks, six heads in each block, $n\_ctx = 1000$, and $embedding\_dim = 300$. We borrow other hyperparameters from \cite{radford2019language-gpt2}.
We limit the token vocabulary size to 100k, which covers over 90\% of the tokens used in the training dataset.
For \RootPath, we limit the maximum length of the path from leaf node to root to be 13, which covers over 90\% of the nodes. For any path longer than 13, we keep the nodes closest to the leaf, and truncate the nodes near the root.

In our implementation, we did not use positional encoding~\cite{vaswani2017attention} or positional embedding~\cite{radford2019language-gpt2} to provide extra positional information over elements
since our early trials with \LeafSeq suggested 
positional embedding is rather hurting than helping. 
This is also supported by the claims in \cite{Irie_2019} that positional encoding does not help for language modeling. 
Recently, \cite{shiv2019novel} tried to introduce tree structures to Transformer models via positional encoding. However, their relative improvement is small compared to what we see with tree-relational prior in Section~\ref{sec:evaluation}.

\paragraph{RNN} For \SrcRNN, we adapt the PyTorch LSTM example implementation~\footnote{\url{https://github.com/pytorch/examples/tree/master/word_language_model}}. We use embedding dimension $d_\model = 300$, with $dropout = 0.5$ and $n\_layers = 1$. We maintain the same vocabulary size at 100k.

\paragraph{\CSeq} For \CSeq, we used a PyTorch adaptation of the publicly released model\footnote{https://github.com/tech-srl/code2seq}, using the same hyperparameters, except changing the vocab size to 100k. For selecting 200 (max number of paths) paths per AST, we first picked paths that ended with the target (to maximize the amount of local context). 
Since for each prediction point in the AST, a new set of leaf-to-leaf paths have to be generated, the data processing for \CSeq takes a substantial amount of time (magnitude of days worth of time). 

We trained all models on Nvidia Tesla V100 (using 4 GPUs at a time) until the loss converged, with all of the parameters randomly initialized. We used the Adam optimizer with the learning rate set to 1e-3. Implementation details regarding number of epochs until convergence, training time (minutes per epoch), inference time (to evaluate over the py150 dataset), and model size, are listed in Table~\ref{tab:impldetails}.

\begin{table*}[ht]
\footnotesize
    \centering
    \begin{tabular}{l|ccc|ccc}
    \hline
                     & \multicolumn{3}{c}{\textbf{Prior work}} & \multicolumn{3}{c}{\textbf{Our work}} \\
        \hline
       & \SrcRNN & \ETH &  \CSeq & \SrcSeq & \RootPath & \DFS    \\
    \hline
      Num epochs        & 9 & n/a & 10 & 9 & 16 & 11    \\
      Training time (min / epoch)        & 45  & n/a & 210  & 45 & 45 	& 60   \\
      Inference time (min)  & 40  & 75 & 45 & 40 & 20 	& 50     \\
      Model size (MB) & 233  & n/a & 149 & 163 & 280	& 163   \\
    \hline
    \end{tabular}
    \caption{Implementation details for all the models - number of epochs until convergence, training time (minutes per epoch), inference time (to evaluate over the py150 dataset), and model size (state dict of PyTorch - the learnable parameters of the model). Note that some information about \ETH is not available since the authors have shared only the model.}
    \label{tab:impldetails}
\end{table*}

\paragraph{Deep3} For the Deep3 model, since the authors have shared only the model and not the training algorithm, we used the model pretrained on py150.

\subsection{Evaluation Metric}
\label{sec:eval-metric}
We evaluate the models on next token prediction for the leaf tokens. We report numbers for all leaf token predictions, as well as breaking down into more interesting categories: attribute access, numeric constant, name (variable, module), and function parameter name. 

To measure performance on these tasks, we use mean reciprocal rank (MRR). 
The rank is defined as 
\begin{equation}
    \textit{MRR} = \frac{1}{n} \sum_{i = 1}^{n} \frac{1}{\textit{rank}_i}
\end{equation}
where $n$ is the number of predicting locations and $rank_i$ is the rank of the correct label given by the model for the $i^{th}$ data point. We present MRR as a percentage, in keeping with prior work~\cite{karampatsis2020big-bpe,hellendoorn2017are-deep-best}.

While Acc@1 only gives score when the correct label is ranked at the top,
MRR also give scores when the true label is not ranked as the top, but among top few prediction. 
Comparing to the hit-or-miss style metric (Acc@1), this is closer to the realistic scenario when completion suggestions are presented to developers.
With this practical perspective and for ease of computation, we only consider $rank_i \le 10$ for each  location $i$ (all $rank_i > 10$ will have a score of 0).  
We share our data processing scripts and model implementations at \url{https://github.com/facebookresearch/code-prediction-transformer}.

\section{Evaluation}
\label{sec:evaluation}


\textbf{RQ1: Given a source token sequence, does a Transformer work better than an RNN, as in previous work?}
Comparing \SrcRNN and \DFS, we find that a Transformer does work better than an RNN. 
For the py150 dataset, we can see a significant improvement in MRR for predicting all leaf tokens in Table~\ref{tab:ideresults-py150}, from 36.6\% to 50.1\% for the \SrcRNN and \SrcSeq models, respectively. The same holds for comparing on the internal dataset, as shown in Table~\ref{tab:ideresults-internal}: 23.8\% vs 36.5\%. Consistent improvements can be seen for specific types of leaf tokens.



\begin{table}[ht]
\footnotesize
    \centering
    \begin{tabular}{l|c|cc}
    \hline
                     & \multicolumn{1}{c}{\textbf{Prior work}} & \multicolumn{2}{c}{\textbf{Our work}} \\
    \hline
      \textbf{Applications} & \SrcRNN  & \SrcSeq & \makecell{\DFS \\(type + value)} \\
    \hline
      Attribute access        & 39.3\% & 55.9\% & \textbf{60.4\%}  \\
      Numeric constant        & 40.6\% & 55.9\% & \textbf{57.0\%}  \\
      Name (variable, module) & 38.2\% & 54.1\% & \textbf{62.7\%}  \\
      Function parameter name & 57.7\%  & \textbf{66.2\%} & 64.8\%   \\ \hline
      All leaf tokens         & 36.6\%  & 50.1\% &	\textbf{54.9\%} \\
    \hline
    \end{tabular}
    \caption{MRR of various types of next token predictions for py150. To fairly compare these models, \DFS makes two predictions - one for the leaf node and then one for its parent internal node (see RQ3 for details).}
    \label{tab:ideresults-py150} 
\end{table}

\begin{table}[ht]
\footnotesize
    \centering
    \begin{tabular}{l|c|cc}
    \hline
                     & \multicolumn{1}{c}{\textbf{Prior work}} & \multicolumn{2}{c}{\textbf{Our work}} \\
    \hline
      \textbf{Applications} & \SrcRNN  & \SrcSeq & \makecell{\DFS \\(type + value)} \\
    \hline
      Attribute access        & 26.4\% & 41.0\% & \textbf{44.5\%}  \\
      Numeric constant        & 32.2\% & 51.7\% & \textbf{53.0\%}  \\
      Name (variable, module) & 25.0\% & 39.3\% & \textbf{47.2\%}  \\
      Function parameter name & 45.5\%  & \textbf{54.3\%} & 51.4\%   \\ \hline
      All leaf tokens         & 23.8\%  & 36.5\% &	\textbf{40.7\%} \\
    \hline
    \end{tabular}
    \caption{MRR of various types of next token predictions for internal dataset. \DFS makes two predictions - one for the leaf node and one for its parent internal node (see RQ3 for details).}
    \label{tab:ideresults-internal}
\end{table}

\textbf{RQ2: Do the Transformer models on tree outperform previous work on AST-based prediction?}

We compare \ETH and \CSeq against \RootPath, and \DFS (\tabref{tab:resultspy150}).
Overall, we found that both transformer-based models, achieve better scores than both \ETH and \CSeq for all leaf tokens as well as for specific types of leaf tokens. Our best performing model, \DFS, improves \ETH's MRR by 14.1\% (from 43.9\% to 58.0\%), and \CSeq's MRR by 14.4\% (from 43.7\% to 58.0\%).
Similar results can be seen for the internal dataset (\tabref{tab:resultsint}).

We also compared the accuracy on AST internal predictions, comparison \ETH and \DFS, as they are the only models with this capability. In Table~\ref{tab:results_types} we see that \DFS improves accuracy over \ETH across the board.
\tabref{tab:results_types} show non-terminal value prediction for both py150 and internal dataset, respectively. We see that \DFS is outperforms \ETH for all internal node types as well.~\footnote{We do not include \CSeq comparison for non-terminal node predictions due to the overhead required to prepare and process the dataset. Since the main part of the paper was on leaf token prediction, and we have shown that \DFS performs significantly better than \CSeq, we did not deem it essential to include the results on non-terminal value predictions.} 

\begin{table}[ht]
\footnotesize
    \centering
    \begin{tabular}{l|cc|cc}
    \hline
                     & \multicolumn{2}{c}{\textbf{Prior work}} & \multicolumn{2}{c}{\textbf{Our work}} \\
    \hline
      \textbf{Applications} & \ETH & \CSeq &\RootPath & \DFS   \\
    \hline
      Attribute access         & 45.3\% & 39.3\%  & 57.2\% & \textbf{60.5\%} \\
      Numeric constant         & 53.2\% & 49.5\% & 59.1\% & \textbf{63.5\%} \\
      Name (variable, module)  & 48.9\% & 45.8\% & 63.5\% & \textbf{66.6\%} \\
      Function parameter name  & 58.1\% & 56.8\% & 65.8\% & \textbf{67.2\%} \\ \hline
      All leaf nodes           & 43.9\% & 43.7\% & 55.1\% & \textbf{58.0\%} \\
    \hline
    \end{tabular}
    \caption{MRR of various types of next token predictions for py150.}
    \label{tab:resultspy150}
\end{table}


\begin{table}[ht]
\footnotesize
    \centering
    \begin{tabular}{l|cc|cc}
    \hline
                     & \multicolumn{2}{c}{\textbf{Prior work}} & \multicolumn{2}{c}{\textbf{Our work}} \\
        \hline
      \textbf{Applications}       & \ETH &  \CSeq  & \RootPath & \DFS    \\
    \hline
      Attribute access         & 38.5\% & 26.4\%  & 41.5\%  & \textbf{44.7\%}   \\
      Numeric constant         & 46.5\% & 45.4\% & 56.1\%  & \textbf{61.5\%}   \\
      Name (variable, module)  & 41.0\% & 31.2\% & 48.0\%  & \textbf{50.7\%}   \\
      Function parameter name  & 50.6\% & 39.3\%  & 52.1\%  & \textbf{53.3\%}   \\ \hline
      All leaf nodes           & 31.6\% & 31.0\% & 40.8\%  & \textbf{43.9\%}   \\
    \hline
    \end{tabular}
    \caption{MRR of various types of next token value prediction for internal dataset.}
    \label{tab:resultsint}
\end{table}



\begin{table}[ht]
\footnotesize
    \centering
    \begin{tabular}{l|c|c|c|c}
    \hline
         \textbf{Dataset}    & \multicolumn{2}{c|}{\textbf{py150}} &  \multicolumn{2}{c}{\textbf{internal}} \\
    \hline
    
      \textbf{Applications}   & Deep3 & \DFS & Deep3 & \DFS    \\
    \hline
      Function call    & 81.6\%  & \textbf{88.5\%}  & 78.2\%  & \textbf{86.0\%} \\
      Assignment       & 76.5\%  & \textbf{78.9\%}  & 78.5\%  & \textbf{79.7\%} \\
      Return           & 52.8\%  & \textbf{67.8\%}  & 59.9\%  & \textbf{72.2\%} \\
      List             & 59.4\%  & \textbf{76.0\%}  & 40.8\%  & \textbf{63.1\%} \\
      Dictionary       & 66.3\%  & \textbf{15.0\%}  & 39.8\%  & \textbf{23.5\%} \\
      Raise            & 35.0\%  & \textbf{63.3\%}  & 33.5\%  & \textbf{59.3\%} \\ \hline
      All types        & 81.9\%  & \textbf{87.3\%}  & 79.9\%  & \textbf{87.7\%} \\
    \hline
    \end{tabular}
    \caption{MRR of various type predictions for py150 and internal dataset.}
    \label{tab:results_types}
\end{table}


\begin{table}
\footnotesize
    \centering
    \begin{tabular}{l|cc}
    \hline
    \textbf{Applications} & \DFS & \TreeRel \\
    \hline
    Attribute access.       & 60.5\%          & \textbf{61.2\%} \\
    Numeric constant        & \textbf{63.5\%} & \textbf{63.5\%} \\
    Name (variable, module) & 66.6\%          & \textbf{67.7\%} \\
    Function parameter name & \textbf{67.2\%} & 67.0\%          \\ \hline
    All leaf nodes          & 58.0\%          & \textbf{58.8\%} \\
    All internal nodes      & 87.3\%          & \textbf{91.9\%} \\
    \hline
    \end{tabular}
    \caption{MRR \DFS compared against its variant \TreeRel that incorporates more tree structure.}
    \label{tab:dfs_alternate}
\end{table}

\textbf{RQ3: Does adding syntactic structure help? 
}
Comparing \SrcSeq and \DFS, we confirm that adding syntactic structure does help.

Comparing \SrcSeq against \DFS on leaf nodes fairly is a bit tricky. 
A source code token we are looking at here contains both syntactic type and lexical value, which translates to two nodes in the AST. For example, the source code token ``2'' is equivalent to an internal node with value ``Num'' and its child leaf node ``2''.
Thus, for a fair comparison, \DFS has to make two predictions -- one for the leaf node and one for its parent internal node. For example, given the sequence ``y ='', and the next prediction should be ``2'', \DFS must predict both the internal "Num" node as well as the leaf "2" node. 
For evaluation, we implemented a local beam search to choose the pair with the maximum joint probability. Table~\ref{tab:ideresults-py150} shows that \DFS still performs better than \SrcRNN and \SrcSeq (except for predicting function parameter name) for predicting the leaf tokens.

It is important to note that even with this correction, 
it is tricky to completely fairly compare the accuracy of \SrcSeq against \DFS model on leaf nodes.
The main issue here is that the contexts used to predict a leaf value is different in the two models, because of different
order of seeing information.  
Consider the case of a code fragment \texttt{x = y + z}, and let us say we need to predict what comes after the ``=''.
In the case of source token based prediction, the predictor has seen ``x ='' and would be expected to produce the
next token (``y'').  In an AST, one would first construct an "Assign" internal node, with the right child to be filled next, and
the next prediction would be actually an interior node "BinOpPlus".  The ``y'' would be predicted as the left child
of BinOpPlus.  In a context in which the AST has been augmented with the BinOpPlus node, the prediction of ``y'' has
more information compared to what a source token model did immediately after the ``=''.  This makes a direct comparison of token predictions difficult. One way to achieve better parity between the models would be to use an in-order traversal instead of a pre-order traversal; we chose the latter because we want to do top-down generation, which will be useful in future AST generative models. 

We also compare \DFS with its variant \TreeRel that extends the model to incorporate even more tree structure, as mentioned in \secref{sec:model-tree}. 
~\tabref{tab:dfs_alternate} shows a slight increase in MRR: 58.0\% vs 58.8\% for all leaf nodes, and 87.3\% to 91.9\% for internal nodes.  Due to mixed benefit on leaf node predictions, we continue to treat \DFS as our flagship model. But this does hint at the existence of more powerful models, if we incorporate more tree structure.

\section{Model Inspection}
\label{sec:inspection}
Despite their wide adoption in many areas of computing, deep learning models have been repeatedly shown susceptible to adversarial examples \cite{akhtar2018threat, chakraborty2018adversarial, zhang2020adversarial}.
Interpretability studies \cite{chakraborty2017interpretability, zhang2018visual, carvalho2019machine, huang2020survey} that provide human-understandable justifications, have thus become an important property in building safe and trustworthy AI systems.
With the growing adoption of neural models for software engineering tasks, the same concern raises.
Recent discoveries of unexpected behaviours of deep learning models of code \cite{wang2019coset, yefet2020adversarial, ramakrishnan2020semantic} calls attention for such studies.
Besides of helping understand the behaviour of deep learning models, interpretability methods have also been used to directly improve the process of software engineering. 
For example, \cite{jiarpakdee2020empirical, wattanakriengkrai2020predicting} used LIME~\cite{ribeiro2016should}, a model-agnostic explainability method, to pinpoint the defective lines of code from defect detection models.


As a crucial first step towards interpretability,
in this section, we reveal what \DFS has learned that leads to its good  predicative power.
We found that \DFS has learned to attribute the prediction to relevant previous tokens. 
While this is not a principled study, we hope it sheds light on a possible approach to perform a complete model inspection study.



Although our Transformer-based models heavily relies on attentions, direct visualizations of weights in individual attention heads did not yield clear insights as the attentions are stacked across layers and multiple attention heads are in effect in each layer (see \secref{sec:trans-intro}). 
We thus turn to gradient-based saliency maps~\cite{simonyan2013deep-saliency, smilkov2017smoothgrad, sundararajan2017axiomatic}, for a more comprehensive account of the influence of each input token.
Following \cite{simonyan2013deep-saliency},
the influence of each input token is computed by taking the partial derivative of the loss function with respect to the embedding vector of that token.
\figref{fig:dfs-pred-influence} is the saliency map we got for \DFS, which visualizes the magnitudes of the gradients fall at each input token ($x$-axis) when the model predicts a particular output ($y$-axis).
Intuitively, the larger the value for a particular token, the more sensitive the output is to the variations at that input token. 

We first observe that the parent AST node (the internal node right above the leaf) is generally important in predicting the next leaf node value: the last token usually has a high influence. 
More generally, we found that \DFS tends to attribute more towards the internal nodes that are directly followed by relevant leaf nodes.
For example, we can see the prediction of \code{atoi} is influenced by (the locations previous to) \code{string}, \code{map} and \code{num\_requests}.
It is not shown in the figure but the predictions of \code{sys} and \code{argv} before \code{2} are influenced by the previous occurrence of the same values followed by \code{0} and \code{1}.
The prediction of \code{host}, \code{gethostbyname} are influenced by (the location previous to) \code{ip}.
For the position of \code{gethostbyname}, \DFS predicts \code{socket} with probability 0.31, with the correct answer ranked second with slightly lower probability 0.24.
All above suggest that \DFS is capable of picking up information from relevant previous code locations. 
The reason \DFS tends to focus more on the parent nodes may be that all relevant information for the next token are already summarized in the hidden representation at the location previous to the token (aka parent nodes for the leaf nodes). 

On an orthogonal note, we also observe that for many predicting locations, the magnitude of gradients are very small,
suggesting the robustness of the model in the sense that it is less sensitive to minor perturbations of the input sequence.

\definecolor{htmlgreen}{RGB}{0, 128, 0}
\begin{figure}
    \centering
        \includegraphics[width=0.5\textwidth]{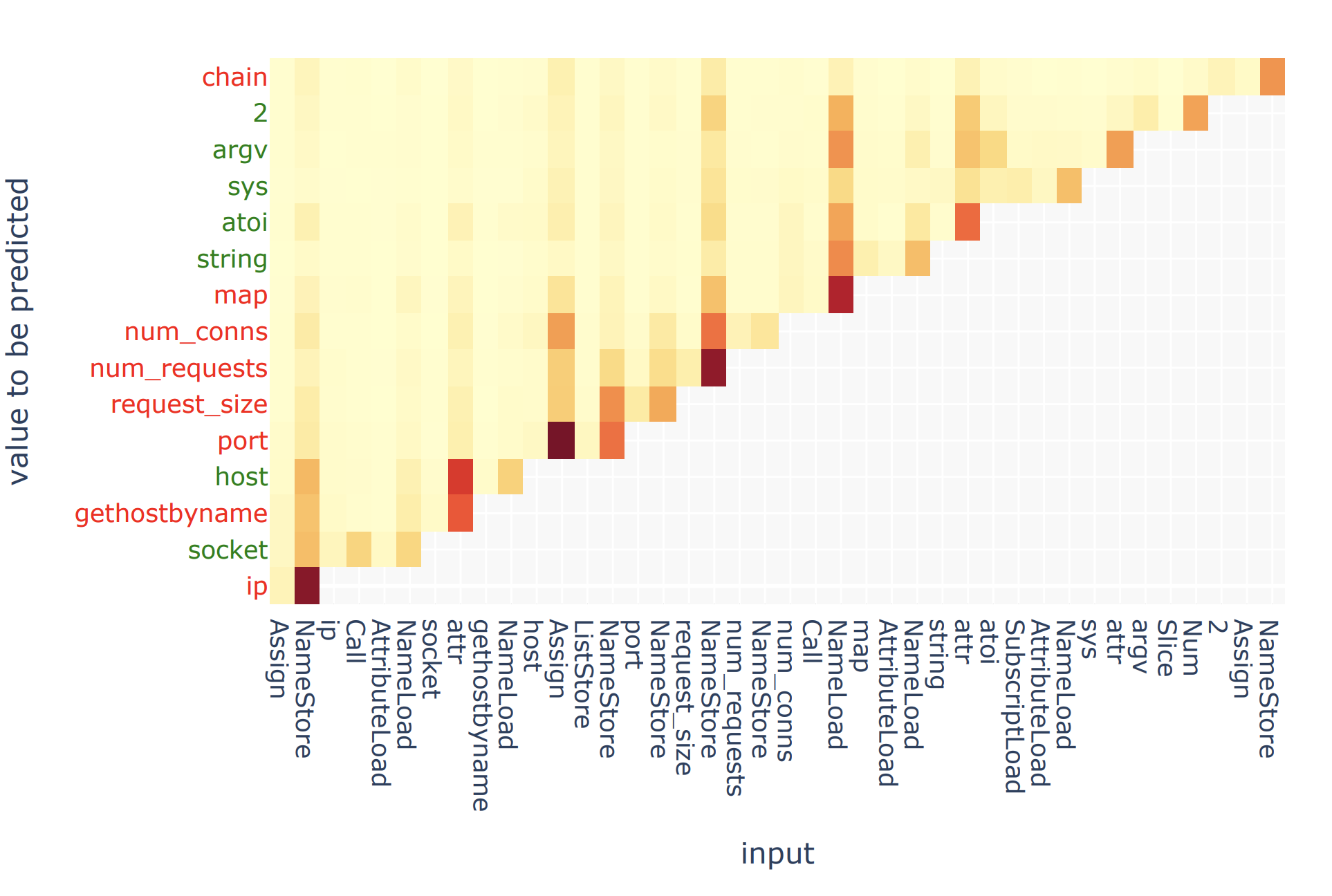}
    \caption{Influence of previous nodes in value prediction of the example in ~\figref{fig:exampleast} by \DFS.
    $x$-axis is labeled with the input values. 
    $y$-axis is labeled with the values to be predicted. 
    Color indicates the model's prediction is \textcolor{htmlgreen}{correct} or \textcolor{red}{wrong}.}
        \label{fig:dfs-pred-influence}
\end{figure}

\section{Threats to Validity}
\label{sec:threats}

\paragraph{Out-of-vocabulary (OOV) handling}
\label{sec:threats/oov}

One of the challenges in next-token prediction is predicting tokens that are rare or unseen in the training corpus.  

When trained on py150, 21\% and 35\% of prediction locations observe OOV tokens for the py150 and the internal test set, respectively.
None of the models in our main evaluation handle OOV predictions and thus are unable to correctly predict for such locations.

Handling of OOV tokens, while important, is orthogonal to core model design, and is outside the scope of our current work.  
Adding OOV handling to a model should further improve its performance, as suggested by previous works~\cite{allamanis2016convolutional, li2018code-rnn-attn, karampatsis2020big-bpe}.  Moreover, a quick comparison (see \secref{sec:related-works}) against 
\PointerMixture from ~\cite{li2018code-rnn-attn}  found that \DFS already has an edge even without OOV handling for the py150 dataset.
We thus do not expect OOV handling to reverse our findings.



\paragraph{Training Corpus}
While larger Python corpora have appeared,  py150 is still sizable at ~500MB; we do not expect the larger corpora to reverse our findings.

\paragraph{Python specificity}
We have only carried out evaluations on Python, and have not demonstrated that our results would carry over (in trends) to other languages.  The Deep3 paper did find their results (in trends) to roughly carry over from Python to JavaScript. 

\paragraph{Model comparison}
To bound the scope of this paper, we limit our investigation to techniques that \emph{do not} require any compiler support beyond constructing the AST; thus, we exclude ways to communicate program analysis results such as def-use information, as in \cite{allamanis2018learning-graph}.
We also limit our study to the Transformer architecture, and purposely exclude the graph neural networks (GNN), because there is already an active debate between GNN and Transformer for NLP applications that transcends code prediction.

\begin{table*}[ht]
\footnotesize
    \centering
    \begin{tabular}{l|c|c|c|c|c|c}
    \hline
         \textbf{Dataset}    & \multicolumn{3}{c|}{\textbf{py150}} &  \multicolumn{3}{c}{\textbf{internal}} \\
    \hline
      \textbf{Applications}   & \PointerMixture & \DFS & OOV Rate (\%) & \PointerMixture & \DFS & OOV Rate (\%)   \\
    \hline
      Attribute access        & 54.2\%  & \textbf{60.5\%} & 19.3\% & \textbf{50.4\%}  & 44.7\% & 33.8\% \\
      Numeric constant        & 50.6\%  & \textbf{63.5\%} & 10.3\% & 40.4\%  & \textbf{61.5\%} & 6.3\% \\
      Name (variable, module) & 49.5\%  & \textbf{66.6\%} & 15.5\% & 42.1\%  & \textbf{50.7\%} &31.4\% \\
      Function parameter name & 59.7\%  & \textbf{67.2\%} & 7.7\% & \textbf{55.2\%}  & 53.3\% & 18.7\% \\ \hline
      All leaf nodes          & 51.9\%  & \textbf{58.0\%} & 21.1\% & \textbf{46.3\%}  & 43.9\% & 34.8\% \\
    \hline
    \end{tabular}
    \caption{MRR of \PointerMixture 
    compared against \DFS for various types of next token prediction for py150 and internal dataset. The out-of-vocabulary (OOV) rates for internal dataset is much higher compared to py150 dataset. }
    \label{tab:pointermixture_results}
\end{table*}


\section{Related Work}
\label{sec:related-works}

We compared our models extensively with three baselines: \SrcRNN, \ETH and \CSeq, both qualitatively in ~\secref{sec:models} and quantitatively in ~\secref{sec:evaluation}.  Before proceeding to a broader discussion, we briefly compare our work with the attentive pointer mixture model (\PointerMixture) from Li at al.~\cite{li2018code-rnn-attn} as another plausible baseline.
We are particularly interested in this model, because it
(1) works with serialized AST node sequences, (2) uses {attention mechanism} %
on top of LSTM, and (3) has reportedly achieved better accuracy than \ETH{}'s results.

The similarities and differences in relation to our work are as follows.
For (1), the preorder traversal of the AST they used is similar to ours. They did not separate types and values into different nodes 
as we do, which makes their sequences more compact than ours.
For (2), their attention is applied within a fixed-sized window on top of LSTM outputs, whereas our Transformer-based models use stacked layers of attentions as the key mechanism for computing throughout the network. 
\cite{li2018code-rnn-attn} also used ``parent attention'', which is based on the parent-child relation in the AST. 
In comparison, our \RootPath and \TreeRel extend far beyond the direct parent-child relation by making use of paths.

For (3), the accuracy reported in \cite{li2018code-rnn-attn} included the predictions of null values for internal nodes, whereas our numbers only consider leaf nodes with non-empty values. 
To fairly compare the predicative power between our proposal and theirs, we implemented \PointerMixture in our setting.%
\footnote{%
To maintain consistency, the hidden size and the embedding sizes were set to 300, and the vocab size was increased to 100k. 
The model was trained for 8 epochs. 
}
Note that \PointerMixture uses a \emph{pointer network}~\cite{vinyals2015pointer} which decides for each point of prediction whether to use the LSTM output or to {copy} from a previous location, whereas our \DFS has no means for handling out-of-vocabulary (OOV) tokens (i.e. all predictions requiring an OOV token are incorrect.)%

We found on the basis of py150 dataset that \PointerMixture outperforms \SrcRNN, as well as the other baselines of \ETH and \CSeq. However, \DFS \textit{outperforms} \PointerMixture, even though all the OOV predictions count as wrong for \DFS!
For the internal dataset, \PointerMixture model does better than \DFS for some prediction types. 
This is expected: we trained the model for illustration on a different dataset (py150) than the one for evaluation (internal), causing a significantly higher percentage of OOV predictions. In actual usage, we would retrain our models on the internal dataset.
The results are shown in \tabref{tab:pointermixture_results}, focusing on \PointerMixture vs. \DFS.

Since we have already introduced a brief history of autocomplete in the paper, this section will focus on other explorations in the Transformer and code prediction space.

\paragraph{Code Prediction}
In this paper, we viewed the context of prediction as code that appears strictly before the cursor~\cite{allamanis2014mining, bielik2016phog, raychev2016learning-noisy, raychev2016probabilistic-deep3-eth-dt, liu2016neural-code-completion, li2018code-rnn-attn, liu2020modeling-stack-lstm}.
Among other flavors of code prediction are the ones where code after the prediction location, when available, is taken into account~\cite{raychev2014code-api,allamanis2018learning-graph,brockschmidt2018generative-graph,alon2020structural-anygen}, e.g. when completing a ``hole'' in a program, or correcting a misused local variable instance.  Another dimension of work
considers prediction at varying granularities of predictions, e.g. from characters~\cite{bielik2016program-character} to subtokens~\cite{karampatsis2020big-bpe})
to AST fragments (e.g. sub-ASTs~\cite{alon2020structural-anygen}). In the context of autocomplete, we believe that subtokens or BPE are indeed a promising future direction, as we mentioned in \secref{sec:future-work}.

There have also been work discussing the practical implications of applying a code prediction tool into production. \cite{hellendoorn_casestudy, ari} state that synthetic benchmarks are not representative of real-world data, and accuracy of the models drops when evaluated on real-world data. \cite{svyatkovskiy2020fast} discusses approaches to make models more lightweight to allow for faster computations and less memory usage in an IDE.  A user evaluation of an autocomplete tool with stronger ML models is outside the scope of this paper.  We  would  also  like  to  point  out  that  advantages  in idealistic settings often transfer to advantages in more practical settings, as confirmed by the results reported in the above two works (Table II and III in \cite{hellendoorn_casestudy}, and Table 2 in \cite{svyatkovskiy2020fast}).

\paragraph{Transformers}
Other than next token prediction, 
Transformers have been used recently for code summarization~\cite{great_iclr20}. Furthermore, there has been a surge of interest since 2019 in extending Transformer models to handle beyond sequential structures for NLP~\cite{wang2019tree-mask,ahmed2019you-traverse,nguyen2020treestructured-hierarchical}.
It has been shown that taking tree structure into account helped code correction~\cite{harer2019tree-correction} and code translation~\cite{shiv2019novel}. 

There is practical interest, outside of academic literature, in the topic of code prediction using Transformers. Galois~\cite{galois} is an open source project that uses GPT-2~\cite{radford2019language-gpt2} for code prediction.
TabNine\texttrademark~published a blog post~\cite{tabnine2019autocompletion} in July 2019 mentioning the use of GPT-2 in their code prediction but revealed no technical detail. Recently a team from Microsoft also published their ongoing efforts~\cite{intellicode} on applying Transformers for code autocomplete. We believe we are among the first to systematically evaluate Transformers for code prediction and compare them to previous models. 

\paragraph{Deep learning techniques over Code, beyond Code Prediction}

There are many other uses of deep learning techniques for code, beyond code prediction.  These include techniques for code summarization~\cite{alon2019code2vec}, bug finding~\cite{allamanis2018learning-graph}, repair~\cite{vasic2019neural} and many others. An interesting aspect of this body of work is in the different ways in which they represent a program as an input to a neural architecture.  These representations have ranged from linear token sequence (as for code prediction~\cite{hellendoorn2017are-deep-best, li2018code-rnn-attn, karampatsis2020big-bpe}), to paths in an AST~\cite{alon2019code2vec,alon2018codeseq,alon2020structural-anygen}, and sometimes even ways to convey static analysis information to the neural network ~\cite{allamanis2018learning-graph, brockschmidt2018generative-graph,yang2019improve, great_iclr20}. 

\section{Conclusion and Future Work}
\label{sec:future-work}

In this paper, we presented ways to using the Transformer for code prediction.  We showed that the Transformer outperforms existing models for code prediction, and when supplied with code's structural information, we are able to get even better predictive power.
Attribution study show that our best model tends to focus on relevant code locations for prediction.  

In the future, one avenue we wish to continue working on is handling out-of-vocabulary words better.
Source code presents a difficulty shared with NLP in handling large vocabularies and rare words. The token/word to be predicted in test data may not appear in the training data. This is even more challenging when predicting identifiers, such as method names, variable names, as developers can come up with arbitrary identifier names. Possible mitigation includes copying mechanism~\cite{allamanis2016convolutional,brockschmidt2018generative-graph,fernandes2018structured} and open-vocabulary models~\cite{cvitkovic2019open,karampatsis2020big-bpe}.

This paper focused on predicting the next token, as it is already a challenging task. In future, we also want to explore predicting multiple tokens at a time, i.e. autocompleting entire expressions.

\newpage
\bibliographystyle{IEEEtran}
\bibliography{IEEEabrv,main}


\end{document}